\documentclass[12pt]{article}

\usepackage{amsmath,amssymb}
\usepackage{amsfonts}
\usepackage{graphicx}
\usepackage{a4wide}

\usepackage[T2A]{fontenc}
\usepackage[cp1251]{inputenc}
\usepackage[russian]{babel}

\begin{document}

\noindent {\it Problems of Information Transmission},\\
\noindent vol. 48, no. 3, pp. 3--23, 2012.

\begin{center} {\bf M. V. Burnashev,
\footnote[1]{The research described in this
publication was made possible in part by the Russian Fund for
Fundamental Research (project number 12-01-00905a).}
H. Yamamoto} \end{center}

\vskip 0.4cm

\begin{center}
{\large\bf
ON RELIABILITY FUNCTION  OF GAUSSIAN CHANNEL WITH
NOISY FEEDBACK: ZERO TRANSMISSION RATE}
\end{center}

{\begin{quotation} \normalsize For information transmission a
discrete time channel with independent \\
additive Gaussian noise is
used. There is also feedback channel with independent additive
Gaussian noise, and the transmitter observes without delay all
outputs of the forward channel via that feedback channel.
Transmission of nonexponential number of messages is considered
and the achievable decoding error exponent for such a combination
of channels is investigated. It is shown that for any finite noise
in the feedback channel the achievable error exponent is better
than similar error exponent of the no-feedback channel. Method of
transmission/decoding used in the paper strengthens the earlier
method used by authors for BSC. In particular, for small feedback
noise, it allows to get the gain of 23.6\% (instead of 14.3\%
earlier for BSC).
\end{quotation}}

\vskip 0.7cm

\begin{center}
{\large\bf \S\,1. Introduction and main results}
\end{center}

We consider the discrete time channel with independent additive
Gaussian noise, i.e. if $\mbox{\boldmath $x$} = (x_1,\ldots,x_n)$
is the input codeword then the received block $\mbox{\boldmath
$y$} = (y_1,\ldots,y_n)$ is
\begin{equation}\label{defchan1}
y_{i} = x_{i} + \xi_{i}, \qquad i=1,\ldots,n,
\end{equation}
where $\mbox{\boldmath $\xi$} = (\xi_{1},\ldots,\xi_{n})$ are
independent ${\cal N}(0,1)$--Gaussian random variables, i.e.
${\mathbf E} \xi_{i} = 0,\; {\mathbf E} \xi_{i}^2 = 1$. There is also
a noisy feedback channel which allows to the transmitter to observe
(without delay) all outputs of the forward channel
\begin{equation}\label{deffeed1}
z_{i} = y_{i} + \sigma\eta_{i}, \qquad i=1,\ldots,n,
\end{equation}
where $\mbox{\boldmath $\eta$} = (\eta_{1},\ldots,\eta_{n})$ are
independent (and independent of $\mbox{\boldmath $\xi$}$)
${\cal N}(0,1)$--Gaussian random variables, i.e.
${\mathbf E} \eta_{i} = 0,\; {\mathbf E}\eta_{i}^2 = 1$. The value
$\sigma > 0$, characterizing the feedback channel noise intensity, is
given. No coding is used in the feedback channel (i.e. the receiver
simply re-transmits all received outputs to the transmitter). In
other words, the feedback channel is ``passive'' (see Fig. 1).

\newpage

\begin{picture}(50,70)(30,-20)
\put(80,5){\framebox(70,20){Transm.}}
\put(190,5){\framebox(100,20){AWGN}}
\put(190,-30){\framebox(100,20){AWGN}}
\put(350,5){\framebox(70,20){Receiver}}

\put(290,15){\vector(1,0){60}}
\put(150,15){\vector(1,0){40}}
\put(60,-20){\line(1,0){130}}
\put(30,15){\vector(1,0){50}}
\put(320,-20){\vector(-1,0){30}}
\put(320,-20){\line(0,1){35}}
\put(60,-20){\vector(0,1){35}}
\put(420,15){\vector(1,0){30}}

\put(170,20){$x$}
\put(310,20){$y$}
\put(170,-15){$z$}
\end{picture}

\vskip 1.0cm

\begin{center}
{Fig. \,1. Channel model}
\end{center}

We assume that the input block $\mbox{\boldmath $x$}$ satisfies the
constraint
\begin{equation}\label{constr1}
\sum_{i=1}^{n}x_{i}^{2} \leq nA,
\end{equation}
where $A$ is a given constant. We denote by AWGN$(A)$ the channel
(\ref{defchan1}) with constraint (\ref{constr1}) without feedback,
and by AWGN$(A,\sigma)$ that channel with noisy feedback
(\ref{deffeed1}).

Since Shannon's paper \cite{Shannon56} it has been known that even
noiseless feedback does not increase the capacity of the Gaussian
channel (or any other memoryless channel). However, feedback allows
to improve the decoding error probability (or simplify the
effective transmission methods). In the case of noiseless feedback
possibility of such improvement of the decoding error probability
with respect to no-feedback channel was shown for a number of
channels in [2--9].

We consider the case when the overall transmission time $n$ and
$M= e^{o(n)}$ equiprobable messages
$\{\theta_{1},\ldots,\theta_{M}\}$ are given. After the moment
$n$, the receiver makes a decision ${\hat \theta}$ on the message
transmitted. We are interested in the best possible decoding error
exponent (and whether it exceeds the similar exponent of the channel
without feedback).

Such problem (for $R = 0$) was first considered in \cite{{BY0}, BY1}
for a binary symmetrical channel. Later in \cite{BurYam1,BY2}, the
case of positive rates (i.e. $R > 0$) was also investigated. The paper
aim is to get similar (in fact, much stronger) results for a Gaussian
channel.

Some results for channels with noiseless feedback can be found in
[2--9], and for the case of noisy feedback -- in
\cite{DrapSah1, KimLapW} (see also discussion in \cite{BY1}).

In order to compare with this paper results, we remind briefly earlier
results from \cite{BY0}--\cite{BY2}. There the binary symmetrical
channel BSC$(p)$ with similar feedback channel BSC$(p_{1})$ was
considered. It was shown in \cite{BY0}--\cite{BY2} that there exists
a certain critical value $p_{\rm crit}(p,R) > 0$, such that if
$p_{1} < p_{\rm crit}(p,R)$, then it is possible to improve the
decoding error exponent of the no--feedback channel. If, in
particular, both $R$ and $p_{1}$ are small then the gain is 14.3\%.
In order to get such improvement the transmission/decoding method
with one ``switching'' moment was developed and investigated.

The method of papers \cite{BY0,BY1} was applied to Gaussian channel
AWGN$(A,\sigma)$ in \cite{XiKim1} with similar to papers
\cite{BY0,BY1} results (in particular, with the same asymptotic gain
14.3\%).

{\it Remark} 1. The transmission method used in
\cite{BY0}--\cite{BY2}, reduces the problem to testing of two most
probable (at some fixed moment) messages. It was mentioned in
\cite[Remark 1]{BY1} and \cite[Remark 3]{BY2} that such method is
not optimal even for one switching moment.

In the paper, still using one switching moment, we essentially
improve the \\
transmission/decoding method of \cite{BY0}--\cite{BY2}.
We show that for any {\it noise intensity} $\sigma^{2} < \infty$ it
is possible to improve the best decoding error exponent $E(M,A)$
of AWGN$(A)$ channel without feedback.

The transmission/decoding method with one switching moment, giving
such improvement is described in \S\S\,2-3. It strengthens the
method introduced by authors earlier in \cite{BY0}--\cite{BY2}.
Of course, if $\sigma$ is not small then the gain is small, but it
is strongly positive. In other words, in the problem considered
there is no any critical level $\sigma_{\rm crit}$, beyond which
it is not possible to improve the exponent $E(M,A)$.

{\it Remark} 2. The paper methods can be applied for BSC as well,
strengthening the results of \cite{BY0}--\cite{BY2}. In particular,
for BSC there is no critical level $p_{\rm crit}(p) < 1/2$, beyond
which it is not possible to improve the exponent $E(M,p)$.

{\it Remark} 3. We consider the case when feedback noise intensity
$\sigma^{2} > 0$ is fixed and does not depend on the number of
messages $M$. The case when the value $\sigma^{2}M$ is small,
corresponds, in a sense, to the noiseless feedback case
(cf. \cite{XiKim1}).

For $\mbox{\boldmath $x$},\mbox{\boldmath $y$} \in
{\mathbb{R}}^{n}$ denote
$$
(\mbox{\boldmath $x$},\mbox{\boldmath $y$}) =
\sum\limits_{i=1}^n x_{i} y_{i}, \quad \|\mbox{\boldmath $x$}\|^2 =
(\mbox{\boldmath $x$},\mbox{\boldmath $x$}), \quad
d\left(\mbox{\boldmath $x$},\mbox{\boldmath $y$}\right) =
\|\mbox{\boldmath $x$} - \mbox{\boldmath $y$}\|^{2}.
$$
A subset ${\cal C} = \{\mbox{\boldmath $x$}_{1},
\ldots, \mbox{\boldmath $x$}_{M}\}$ with
$\|\mbox{\boldmath $x$}_{i}\|^{2} = An$, $i = 1,\ldots,M$ is
called a $(M,A,n)$--code of length $n$.

For a code ${\cal C} = \{\mbox{\boldmath $x$}_{i}\}$ denote by
$P_{\rm e}({\cal C})$ the minimal possible decoding error
probability
$$
P_{\rm e}({\cal C}) = \min \max_{i}P(e|\mbox{\boldmath $x$}_{i}),
$$
where $P(e|\mbox{\boldmath $x$}_{i})$ -- conditional decoding error
probability provided $\mbox{\boldmath $x$}_{i}$ was transmitted,
and minimum is taken over all decoding methods
(it will be convenient for us to denote the transmitted message
both $\theta_{i}$ and $\mbox{\boldmath $x$}_{i}$).

For $M$ messages and AWGN$(A)$ channel denote by $P_{\rm e}(M,A,n)$
the minimal possible decoding error probability for the best
$(M,A,n)$--code. We are interested in the best exponent (in $n$) of
that function
$$
E(M,A) = \limsup_{n \to \infty}\,\frac{1}{n}\,
\ln \frac{1}{P_{\rm e}(M,A,n)}.
$$

Similarly, for AWGN$(A,\sigma)$ channel denote by
$P_{\rm e}(M,A,\sigma,n)$ the minimal possible \\
decoding error probability and introduce the function
$$
F(M,A,\sigma) = \limsup_{n \to \infty}\,\frac{1}{n}\, \ln
\frac{1}{P_{\rm e}(M,A,\sigma,n)}.
$$

In the paper we consider the case when $M$ is a fixed number of
messages, or $M = M_{n} \to \infty$ as $n \to \infty$, but
$M_{n} = e^{o(n)}$ (it corresponds to zero-rate of transmission).

It is known that $E(M,A)$ is attained for a simplex code
\cite{Sh, Pin1}
\begin{equation}\label{main2}
\begin{gathered}
E(M,A) = \frac{AM}{4(M-1)}.
\end{gathered}
\end{equation}

It is also known that if $\sigma = 0$ (i.e. in the case of noiseless
feedback) then for a fixed $M$ \cite{Pin1}
$$
\begin{gathered}
F(M,A,0) =  \frac{A}{2}.
\end{gathered}
$$

For AWGN$(A,\sigma)$ channel denote by $F_{1}(M,A,\sigma)$ the best
error exponent for the \\
transmission method with one switching moment, described in \S\S 2--3.
Then $F_{1}(M,A,\sigma) \leq F(M,A,\sigma)$ for all $M,A,\sigma$.

One of two the paper main results is as follows.

{T h e o r e m \,1}. Let $\ln M = o(n)$, $n \to \infty$. Then:

a) {\it If $\sigma \to 0$ then the formula holds
\begin{equation}\label{Theor1a}
\begin{gathered}
F_{1}(M,A,\sigma) \geq \frac{AM}{4(M-1)}\left[1 +
\frac{1}{2+\sqrt{5}}- \frac{1}{2M} + o(1)\right].
\end{gathered}
\end{equation}
Since $1/(2+\sqrt{5}) \approx 0.236$, then for large $M$ the
formula} (\ref{Theor1a}) {\it gives} 23.6\% {\it of improvement with
respect to no-feedback channel}.

b) {\it If $\sigma \to \infty$ then the formula holds}
\begin{equation}\label{Theor1b}
\begin{gathered}
F_{1}(M,A,\sigma) \geq \frac{AM}{4(M-1)}\left[1 +
\frac{1}{56\sigma^{2}} + O(\sigma^{-4})\right] > E(M,A) =
\frac{AM}{4(M-1)}.
\end{gathered}
\end{equation}

In \S 3 the second paper main results -- a more general theorem 2,
valid for any $\sigma^{2} < \infty$, is proved. Theorem 1 follows
from it.

In a standard way reliability functions $E(R,A)$ and $F(R,A,\sigma)$
of no-feedback channel and AWGN$(A,\sigma)$ channel with noisy
feedback can be defined. Then from theorem 1 we get

{C o r o l l a r y}. a) For $\sigma \to 0$ and $R = 0$ the formula
holds
\begin{equation}\label{Cor1a}
\begin{gathered}
F(0,A,\sigma) \geq F_{1}(0,A,\sigma) \geq \frac{A}{4}\left[1 +
\frac{1}{2+\sqrt{5}}+ o(1)\right].
\end{gathered}
\end{equation}

b) {\it For $\sigma \to \infty$ and $R = 0$ the inequality holds}
\begin{equation}\label{Cor1b}
\begin{gathered}
F(0,A,\sigma) \geq F_{1}(0,A,\sigma) \geq \frac{A}{4}\left[1 +
\frac{1}{56\sigma^{2}} + O(\sigma^{-4})\right] > E(0,A) = \frac{A}{4}.
\end{gathered}
\end{equation}

In order to simplify formulas we will pay attention only to
exponential (in $n$) terms, omitting power factors. Moreover,
$f \sim g$ means that $n^{-1}\ln f = n^{-1}\ln g +
o(1),\,n \to \infty$. Similarly $f \lesssim g$, etc. is meant.
Greek letters $\xi,\eta,\zeta,\xi_{1},\ldots$ designate
${\cal N}(0,1)$--Gaussian random variables.

In \S\,2 the transmission method with one switching moment and in
\S\,3 its decoding are described. In \S\,4 that method is investigated
and general theorem 2 is proved. Using theorem 2 in \S\,5 theorem 1
is proved.

Some preliminary (and simplified) version of the paper results
(without detailed proofs) were published in \cite{BurYam2}.

\begin{center}
{\large\bf \S\,2. Improved transmission/decoding method}
\end{center}

We use the transmission strategy with one fixed switching moment at
which the coding function will be changed. The transmission method
used earlier in \cite{BY0}--\cite{BY2} (and in \cite{XiKim1})
reduced the problem to testing of two most probable (at some fixed
moment) messages. We improve that strategy in both transmission and
decoding stages.

In order to simplify formulas we start with case $M \leq (n+2)/2$.
We partition the total transmission time $[1,n]$ on two phases:
$[1,M-1]$ (phase I) and $[M,2M-2]$ (phase II).
Thus the total length of the code used is $2M -2$. The remaining
time $[2M-1,n]$ is not used. After moment $2M -2$ the receiver makes
a decision in favor of the most probable message $\theta_{i}$ (based
on all received on $[1,2M-2]$ signals).

Each of $M$ codewords $\{\mbox{\boldmath $x$}_{i}\}$ of length
$2M-2$ have the form $\mbox{\boldmath $x$}_{i} =
(\mbox{\boldmath $x$}_{i}',\mbox{\boldmath $x$}_{i}'')$, where
$\mbox{\boldmath $x$}_{i}'$ has length $M-1$ (to be used on phase I)
and $\mbox{\boldmath $x$}_{i}''$ has length $M-1$ (to be used on
phase II). Similarly, the received block $\mbox{\boldmath $y$}$
has the form $\mbox{\boldmath $y$} =
(\mbox{\boldmath $y$}',\mbox{\boldmath $y$}'')$, where
$\mbox{\boldmath $y$}'$ is the block received on phase I and
$\mbox{\boldmath $y$}''$ is the block received on phase II. Denote
by $\mbox{\boldmath $z$}'$ the received (by the transmitter) block
on phase I. The codewords first parts
$\{\mbox{\boldmath $x$}_{i}'\}$ are fixed, while the second parts
$\{\mbox{\boldmath $x$}_{i}''\}$ will depend on the block
$\mbox{\boldmath $z$}'$ received by the transmitter on phase I.

We set two positive constants $A_{1}, A_{2}$ such that
\begin{equation}\label{A1A2II}
A_{1} + A_{2} = nA,
\end{equation}
and denote
\begin{equation}\label{defbeta}
\beta = \frac{A_{2}}{A_{1}}, \qquad A_{3} = \frac{MA_{1}}{M-1},
\qquad A_{4} = \frac{MA_{2}}{M-1}, \qquad
\mu = \frac{A_{2}}{A_{3}} = \frac{(M-1)\beta}{M}.
\end{equation}
Then $A = (1+\beta)A_{1}/n$.

Denoting
$$
d_{i}= d(\mbox{\boldmath $x$}_{i}',\mbox{\boldmath $y$}') =
\|\mbox{\boldmath $y$}' - \mbox{\boldmath $x$}_{i}'\|^{2},
$$
arrange the distances $\{d_{i},\, i=1,\ldots,M\}$  for the
receiver after phase I in the increasing order, and denote
$$
d^{(1)} = \min_{i} d(\mbox{\boldmath $x$}_{i}',
\mbox{\boldmath $y$}') \leq  d^{(2)} \leq \ldots \leq d^{(M)} =
\max_{i} d(\mbox{\boldmath $x$}_{i}',\mbox{\boldmath $y$}')
$$
(case of tie has zero probability). Let also
${\mbox{\boldmath $x$}'}^{(1)},\ldots,{\mbox{\boldmath $x$}'}^{(M)}$
be the corresponding ranking of codewords
$\{{\mbox{\boldmath $x$}'}\}$ after phase I for the receiver, i.e
${\mbox{\boldmath $x$}'}^{(1)}$ is the closest to
$\mbox{\boldmath $y$}'$ codeword, etc.

Similarly, denoting
$$
d_{i}^{(t)}= d(\mbox{\boldmath $x$}_{i}',\mbox{\boldmath $z$}') =
\|\mbox{\boldmath $y$}' - \mbox{\boldmath $z$}_{i}'\|^{2},
$$
arrange the distances $\{d_{i}^{(t)},\,i=1,\ldots,M\}$
for the transmitter after phase I in the increasing
order, and denote
$$
d^{(1)t} = \min_{i} d_{i}^{(t)}  \leq  d^{(2)t} \leq \ldots \leq
d^{(M)t} = \max_{i} d_{i}^{(t)}.
$$
Let also ${\mbox{\boldmath $x$}'}^{(1)t},\ldots,
{\mbox{\boldmath $x$}'}^{(M)t}$ be the corresponding ranking of
codewords $\{{\mbox{\boldmath $x$}'}\}$ after phase I for the
transmitter, i.e ${\mbox{\boldmath $x$}'}^{(1)}$ is the closest to
$\mbox{\boldmath $z$}'$ codeword, etc.

{\bf Transmission}. On phase I the transmitter uses a simplex code
of $M$ codewords $\{\mbox{\boldmath $x$}_{i}'\}$ of length $M-1$
such that $\|\mbox{\boldmath $x$}_{i}'\|^{2} = A_{1}$.

For phase II we set a number $\tau_{0} > 0$. Based on the
received block $\mbox{\boldmath $z$}'$ the transmitter selects three
most probable codewords ${\mbox{\boldmath $x$}'}^{(1)t},
{\mbox{\boldmath $x$}'}^{(2)t}, {\mbox{\boldmath $x$}'}^{(3)t}$ and
calculates for them the value
$$
d_{23}^{(t)} = d^{(3)t} - d^{(2)t} = \tau A_{3} \geq 0.
$$
The code $\{\mbox{\boldmath $x$}_{k}''\}$ with
$\|\mbox{\boldmath $x$}_{k}''\|^{2} = A_{2},\,k=1,\ldots,M$ used by
the transmitter on phase II depends on codewords
${\mbox{\boldmath $x$}'}^{(1)t}, {\mbox{\boldmath $x$}'}^{(2)t},
{\mbox{\boldmath $x$}'}^{(3)t}$ and the value $\tau$ as follows.

C a s e \,1. If after phase I
\begin{equation}\label{1par1r}
d_{23}^{(t)} = \tau A_{3} \leq \tau_{0}A_{3},
\end{equation}
then on phase II the transmitter uses the same simplex code of
$M$ codewords $\{\mbox{\boldmath $x$}_{i}'\}$ of length $M-1$, such
that $\|\mbox{\boldmath $x$}_{i}''\|^{2} = A_{2}$.

C a s e \,2. If after phase I
\begin{equation}\label{1par1ar}
d_{23}^{(t)} = \tau A_{3} > \tau_{0}A_{3},
\end{equation}
then on phase II the transmitter uses another code
$\{\mbox{\boldmath $x$}_{k}''\}$ with
$\|\mbox{\boldmath $x$}_{k}''\|^{2} = A_{2},\,k=1,\ldots,M$:

a) two most probable messages $\theta_{i},\theta_{j}$ have opposite
codewords $\mbox{\boldmath $x$}_{i}'' = -\mbox{\boldmath $x$}_{j}''$
which have nonzero coordinates only at moment $M-2$;

b) remaining $M-2$ messages $\{\theta_{k}\}$ use a simplex code of
$M-2$ codewords $\{\mbox{\boldmath $x$}_{k}''\}$ of length $M-3$
trailed by $0$ at moment $M-2$. All those codewords
$\{\mbox{\boldmath $x$}_{k}''\}$ are orthogonal to the first two
codewords $(\mbox{\boldmath $x$}_{i}'',\mbox{\boldmath $x$}_{j}'')$.

This transmission method strengthens the method used in
\cite{BY0}--\cite{BY2}. The code used in case I helps in the case
when after phase I three most probable codewords
${\mbox{\boldmath $x$}'}^{(1)t},
{\mbox{\boldmath $x$}'}^{(2)t}, {\mbox{\boldmath $x$}'}^{(3)t}$ are
approximately equiprobable.

{\bf Decoding}. Due to noise in the feedback channel the receiver
does not know exactly codewords ${\mbox{\boldmath $x$}'}^{(1)t},
{\mbox{\boldmath $x$}'}^{(2)t}, {\mbox{\boldmath $x$}'}^{(3)t}$ and
the value $\tau$ for them, and therefore it does not know
the code used on phase II. But it may evaluate probabilities of all
possible codewords ${\mbox{\boldmath $x$}'}^{(1)t},
{\mbox{\boldmath $x$}'}^{(2)t}, {\mbox{\boldmath $x$}'}^{(3)t}$ and
the value $\tau$ for them, and so find the probabilities with which
any code was used.

It allows to the receiver, based on the received block
$\mbox{\boldmath $y$}$, to find posterior probabilities
$\{p(\mbox{\boldmath $y$}|\mbox{\boldmath $x$}_{i})\}$ and make
decision in favor of most probable message $\theta_{i}$.
Such full decoding is described in details below.

\begin{center}
{\large\bf \S\,3. Full decoding and error probability $P_{\rm e}$}
\end{center}

Note that
$$
\begin{gathered}
\ln\frac{p\left(\mbox{\boldmath $y$}|
\mbox{\boldmath $x$}_{2}\right)}
{p\left(\mbox{\boldmath $y$}|\mbox{\boldmath $x$}_{1}\right)} =
(\mbox{\boldmath $x$}_{2} - \mbox{\boldmath $x$}_{1},
\mbox{\boldmath $y$}) - \frac{1}{2}\left(
\|\mbox{\boldmath $x$}_{2}\|^{2} -
\|\mbox{\boldmath $x$}_{1}\|^{2}\right).
\end{gathered}
$$
If $\mbox{\boldmath $x$}_{\rm true}$ is the true codeword then
$\mbox{\boldmath $y$} = \mbox{\boldmath $x$}_{\rm true} +
\mbox{\boldmath $\xi$}$ and
$\mbox{\boldmath $\xi$} = (\mbox{\boldmath $\xi$}',
\mbox{\boldmath $\xi$}'') = (\xi_{1},\ldots,\xi_{n})$, where
all $\{\xi_{i}\}$ are independent ${\cal N}(0,1)$--Gaussian random
variables. If
$\mbox{\boldmath $x$}_{\rm true} =  \mbox{\boldmath $x$}_{1}$, then
$$
\ln\frac{p\left(\mbox{\boldmath $y$}|
\mbox{\boldmath $x$}_{2}\right)}
{p\left(\mbox{\boldmath $y$}|\mbox{\boldmath $x$}_{1}\right)} =
(\mbox{\boldmath $x$}_{2} - \mbox{\boldmath $x$}_{1},
\mbox{\boldmath $\xi$}) - \frac{1}{2}
\|\mbox{\boldmath $x$}_{2} - \mbox{\boldmath $x$}_{1}\|^{2},
$$
where $(\mbox{\boldmath $x$},\mbox{\boldmath $\xi$})$ is
${\cal N}(0,\|\mbox{\boldmath $x$}\|^{2})$--Gaussian random
variable.

The receiver makes decision after moment $n$ using all received
block $\mbox{\boldmath $y$}$. If after phase I the difference
$d^{(3)} - d^{(2)}$ is rather close to $\tau_{0}A_{3}$ (see
(\ref{1par1r}) and (\ref{1par1ar})) then due to noise in the
feedback link the receiver can not be sure which code was used by
the transmitter on phase II (since lists
$\{{\mbox{\boldmath $x$}'}^{(1)},{\mbox{\boldmath $x$}'}^{(2)}\}$
and
$\{{\mbox{\boldmath $x$}'}^{(1)t},{\mbox{\boldmath $x$}'}^{(2)t}\}$
may turn out to be different). But based on $\mbox{\boldmath $y$}'$
the receiver knows the probability distribution of the code used by
the transmitter on phase II. Then in the decoding it should take
into account that distribution.

Note that if $\theta_{\rm true} = \theta_{1}$ then
$$
d_{i} - d_{1} = 2A_{3} + 2(\mbox{\boldmath $x$}_{1}' -
\mbox{\boldmath $x$}_{i}',\mbox{\boldmath $\xi$}'), \qquad
i = 2,\ldots,M.
$$

If $\theta_{\rm true} = \theta_{1}$, then for decoding error
probability $P_{\rm e}$ we have
$$
\begin{gathered}
P_{\rm e} = {\mathbf P}\left\{\max_{i \geq 2}
\ln \frac{p\left({\mbox{\boldmath $y$}}\big|\theta_{i}\right)}
{p\left({\mbox{\boldmath $y$}}\big|\theta_{1}\right)} \geq 0
\big|\theta_{1}\right\} \leq M{\mathbf P}\left\{
\ln \frac{p\left({\mbox{\boldmath $y$}}\big|\theta_{2}\right)}
{p\left({\mbox{\boldmath $y$}}\big|\theta_{1}\right)} \geq 0
\big|\theta_{1}\right\} = \\
= M{\mathbf P}\left\{X + Y \geq 0\big|\theta_{1}\right\},
\end{gathered}
$$
where
\begin{equation}\label{XY}
\begin{gathered}
X = \ln\frac{p\left({\mbox{\boldmath $y$}'}\big|\theta_{2}\right)}
{p\left({\mbox{\boldmath $y$}'}\big|\theta_{1}\right)} =
(\mbox{\boldmath $y$}',\mbox{\boldmath $x$}_{2}' -
\mbox{\boldmath $x$}_{1}') = -A_{3} + (\mbox{\boldmath $x$}_{2}' -
\mbox{\boldmath $x$}_{1}',\mbox{\boldmath $\xi$}'), \\
Y = \ln \frac{p\left({\mbox{\boldmath $y$}''}\big|
{\mbox{\boldmath $y$}'},\theta_{2}\right)}
{p\left({\mbox{\boldmath $y$}''}\big|{\mbox{\boldmath $y$}'},
\theta_{1}\right)}.
\end{gathered}
\end{equation}

In order to investigate random variable $Y$ introduce the following
sets of random events (conditions):
$$
\begin{gathered}
{\cal Z}_{1} = \left\{\mbox{\boldmath $z$}': d_{23}^{(t)} \leq
\tau_{0}A_{3}\right\}, \\
{\cal Z}_{2} = \left\{\mbox{\boldmath $z$}': d_{23}^{(t)} >
\tau_{0}A_{3}, \{\mbox{\boldmath $x$}_{1}',
\mbox{\boldmath $x$}_{2}'\} =
\{{\mbox{\boldmath $x$}'}^{(1)t},
{\mbox{\boldmath $x$}'}^{(2)t}\}\right\}, \\
{\cal Z}_{3} = \left\{\mbox{\boldmath $z$}': d_{23}^{(t)} >
\tau_{0}A_{3},\left|\{\mbox{\boldmath $x$}_{1}',
\mbox{\boldmath $x$}_{2}'\}\bigcap \{{\mbox{\boldmath $x$}'}^{(1)t},
{\mbox{\boldmath $x$}'}^{(2)t}\}\right| = 1\right\}, \\
{\cal Z}_{4} = \left\{\mbox{\boldmath $z$}': d_{23}^{(t)} >
\tau_{0}A_{3}, \{\mbox{\boldmath $x$}_{1}',\mbox{\boldmath $x$}_{2}'\}
\bigcap \{{\mbox{\boldmath $x$}'}^{(1)t},
{\mbox{\boldmath $x$}'}^{(2)t}\} = \emptyset\right\}.
\end{gathered}
$$
We assume that the true message is $\theta_{1}$. Then using sets
${\cal Z}_{2},{\cal Z}_{3},{\cal Z}_{4}$ it will be possible to
describe all possible relations between pairs
$\{{\mbox{\boldmath $x$}'}^{(1)},{\mbox{\boldmath $x$}'}^{(2)}\}$ and
$\{{\mbox{\boldmath $x$}'}^{(1)t},{\mbox{\boldmath $x$}'}^{(2)t}\}$
of most probable messages for the receiver and the transmitter,
respectively.

Denote
$$
p_{k} = {\mathbf P}({\cal Z}_{k}\big|\mbox{\boldmath $y$}',
\mbox{\boldmath $x$}_{1}'), \qquad k = 1,\ldots,4.
$$

We have
$$
\begin{gathered}
\frac{p\left(\mbox{\boldmath $y$}''\big|\mbox{\boldmath $y$}',
\theta_{2}\right)}{p\left(\mbox{\boldmath $y$}''\big|
\mbox{\boldmath $y$}',\theta_{1}\right)} =
{\mathbf E}_{\mbox{\boldmath $z$}'|\mbox{\boldmath $y$}'}
\frac{p\left(\mbox{\boldmath $y$}''\big|\mbox{\boldmath $z$}',
\mbox{\boldmath $x$}_{2}''\right)}{p\left(\mbox{\boldmath $y$}''
\big|\mbox{\boldmath $z$}', \mbox{\boldmath $x$}_{1}''
\right)} = {\mathbf E}_{\mbox{\boldmath $z$}'|\mbox{\boldmath $y$}'}
e^{(\mbox{\boldmath $y$}'',\mbox{\boldmath $x$}_{2}'' -
\mbox{\boldmath $x$}_{1}'')} = \sum\limits_{k=1}^{4}p_{k}
e^{(\mbox{\boldmath $y$}'',\mbox{\boldmath $x$}_{2}'' -
\mbox{\boldmath $x$}_{1}'')},
\end{gathered}
$$
where blocks $\mbox{\boldmath $x$}_{1}'',\mbox{\boldmath $x$}_{2}''$
depend on $k$ (via ${\cal Z}_{k}$).
If $\theta_{\rm true} = \theta_{1}$, then $\mbox{\boldmath $y$}'' =
\mbox{\boldmath $x$}_{1}'' + \mbox{\boldmath $\xi$}''$, and
$$
\begin{gathered}
e^{Y} = \frac{p\left({\mbox{\boldmath $y$}''}\big|
{\mbox{\boldmath $y$}'},\theta_{2}\right)}
{p\left({\mbox{\boldmath $y$}''}\big|
{\mbox{\boldmath $y$}'},\theta_{1}\right)} = e^{-A_{2}}
\sum\limits_{k=1}^{4}p_{k}e^{(\mbox{\boldmath $x$}_{1}'',
\mbox{\boldmath $x$}_{2}'') + (\mbox{\boldmath $\xi$}'',
\mbox{\boldmath $x$}_{2}'' - \mbox{\boldmath $x$}_{1}'')}, \\
Y \leq \ln 4 - A_{2} +
\max_{k}\{(\mbox{\boldmath $x$}_{1}'',\mbox{\boldmath $x$}_{2}'') +
(\mbox{\boldmath $\xi$}'',\mbox{\boldmath $x$}_{2}'' -
\mbox{\boldmath $x$}_{1}'') + \ln p_{k}\}.
\end{gathered}
$$
Therefore
$$
\begin{gathered}
P_{\rm e}\leq M{\mathbf P}\left\{X + Y \geq 0\big|\theta_{1}\right\}
= M{\mathbf E}_{\mbox{\boldmath $y$}'}{\mathbf P}\left\{X + Y
\geq 0\big|\mbox{\boldmath $y$}',\theta_{1}\right\} \leq \\
\leq M{\mathbf E}_{\mbox{\boldmath $y$}'}{\mathbf P}
\left\{X + \ln 4 - A_{2} +
\max_{k}\{(\mbox{\boldmath $x$}_{1}'',\mbox{\boldmath $x$}_{2}'') +
(\mbox{\boldmath $\xi$}'',\mbox{\boldmath $x$}_{2}'' -
\mbox{\boldmath $x$}_{1}'') + \ln p_{k}\} \geq 0
\big|\mbox{\boldmath $y$}',\theta_{1}\right\} \leq \\
\leq M\sum_{k=1}^{4}{\mathbf E}_{\mbox{\boldmath $y$}'}
{\mathbf P}\left\{X + \ln 4- A_{2} +
(\mbox{\boldmath $x$}_{1}'',\mbox{\boldmath $x$}_{2}'') +
(\mbox{\boldmath $\xi$}'',\mbox{\boldmath $x$}_{2}'' -
\mbox{\boldmath $x$}_{1}'') + \ln p_{k} \geq 0\big|
\mbox{\boldmath $y$}',{\cal Z}_{k},\theta_{1}\right\} \lesssim \\
\lesssim \frac{1}{2}M\sum_{k=1}^{4}
{\mathbf E}_{\mbox{\boldmath $\xi$}'}
\exp\left\{-\frac{\left[A_{2} - X -
(\mbox{\boldmath $x$}_{1}'',\mbox{\boldmath $x$}_{2}'') -
\ln p_{k}\right]_{+}^{2}}
{2\|\mbox{\boldmath $x$}_{2}'' -\mbox{\boldmath $x$}_{1}''\|^{2}}
\right\} = \frac{1}{2}M\sum_{k=1}^{4}e^{-B_{k}},
\end{gathered}
$$
where
$$
\begin{gathered}
B_{k} = -\ln {\mathbf E}_{\mbox{\boldmath $\xi$}'}e^{-b_{k}},
\quad k =1,\ldots,4,  \\
b_{k} = \frac{\left[A_{2} + A_{3} - (\mbox{\boldmath $x$}_{2}' -
\mbox{\boldmath $x$}_{1}',\mbox{\boldmath $\xi$}') -
(\mbox{\boldmath $x$}_{1}'',\mbox{\boldmath $x$}_{2}'') -
\ln p_{k}\right]_{+}^{2}}{2\|\mbox{\boldmath $x$}_{2}'' -
\mbox{\boldmath $x$}_{1}''\|^{2}}.
\end{gathered}
$$
Here $(\mbox{\boldmath $x$}_{1}' - \mbox{\boldmath $x$}_{2}',
\mbox{\boldmath $\xi$}') = \sqrt{2A_{3}}\,\xi_{i}$, where
$\xi \sim {\cal N}(0,1)$, for all $k$, and
$\mbox{\boldmath $x$}_{1}'',\mbox{\boldmath $x$}_{2}''$
depend on $k$. In particular,
$$
\begin{gathered}
b_{1} = \frac{1}{4A_{4}}\left[A_{3}+A_{4} -
(\mbox{\boldmath $x$}_{2}' - \mbox{\boldmath $x$}_{1}',
\mbox{\boldmath $\xi$}') - \ln p_{1}\right]_{+}^{2}, \\
b_{2} = \frac{1}{8A_{2}}\left[A_{3}+2A_{2} -
(\mbox{\boldmath $x$}_{2}' -\mbox{\boldmath $x$}_{1}',
\mbox{\boldmath $\xi$}') - \ln p_{2}\right]_{+}^{2}, \\
b_{3} = \frac{1}{4A_{2}}\left[A_{3}+A_{2} -
(\mbox{\boldmath $x$}_{2}' - \mbox{\boldmath $x$}_{1}',
\mbox{\boldmath $\xi$}') - \ln p_{3}\right]_{+}^{2}, \\
b_{4} = \frac{(M-3)}{4A_{2}(M-2)}\left[A_{3}+ \frac{(M-2)A_{2}}{M-3}
- (\mbox{\boldmath $x$}_{2}' - \mbox{\boldmath $x$}_{1}',
\mbox{\boldmath $\xi$}') - \ln p_{4}\right]_{+}^{2}.
\end{gathered}
$$
Then
\begin{equation}\label{genFB}
F_{1}(M,A,\sigma) \geq \min\limits_{k=1,\ldots,4}B_{k}.
\end{equation}

We should find values $B_{k},p_{k},\,k=1,\ldots,4$ and choose
optimal parameters $\beta, \tau_{0}$. We show below that in
interesting for us cases probabilities $p_{1},p_{3},p_{4}$ are small,
and therefore the probability $p_{2}$ is close to $1$. Moreover, we
omit estimates for values $p_{4},B_{4}$, since clearly
$p_{4} < p_{3}$ and $B_{4} \geq B_{3}$.

We start with the simplest term $B_{2}$. Note that if
$\xi \sim {\cal N}(0,1)$, then
$$
{\mathbf E}e^{-a(b-\xi)^{2}/2} =
\frac{e^{-ab^{2}/(2+2a)}}{\sqrt{1+a}}, \qquad a > -1.
$$
Neglecting $\ln p_{k}$, we get (as $n \to \infty$)
$$
\begin{gathered}
{\mathbf E}_{\mbox{\boldmath $\xi$}'}e^{-b_{2}} \leq
{\mathbf P}\{b_{2} = 0\} + {\mathbf E}\exp\left\{
-\frac{1}{8A_{2}}\left[A_{3}+2A_{2} -
(\mbox{\boldmath $x$}_{2}' -\mbox{\boldmath $x$}_{1}',
\mbox{\boldmath $\xi$}')\right]^{2}\right\} = \\
= {\mathbf P}\left\{\sqrt{2A_{3}}\,\xi \geq A_{3} + 2A_{2}\right\} +
{\mathbf E}\exp\left\{-
\frac{A_{3}}{4A_{2}}\left[\frac{A_{3}+2A_{2}}{\sqrt{2A_{3}}} - \xi
\right]^{2}\right\} \leq \\
\leq \Phi\left(-\frac{A_{3}+2A_{2}}{\sqrt{2A_{3}}}\right) +
\exp\left\{-\frac{A_{3}+2A_{2}}{4}\right\} \leq \\
\leq 2\exp\left\{-\frac{A_{3}+2A_{2}}{4}\right\} =
2\exp\left\{-\frac{MAn(1+2\mu)}{4(M-1)(1+\beta)}\right\},
\end{gathered}
$$
where we used simple inequality
\begin{equation}\label{Phi1}
\Phi(-z) = \frac{1}{\sqrt{2\pi}}\int\limits_{z}^{\infty}
e^{-u^{2}/2}du \leq \frac{1}{2} e^{-z^{2}/2}, \qquad z \geq 0.
\end{equation}
Inequality (\ref{Phi1}) will be regularly used in the paper.
Therefore
\begin{equation}\label{defe2}
\begin{gathered}
B_{2} \geq \frac{MA(1+2\mu)}{4(M-1)(1+\beta)} - \frac{1}{n}.
\end{gathered}
\end{equation}

Calculation of values $B_{1},B_{3}$ will demand more efforts. It is
done in the next section.

\begin{center}
{\large\bf \S\,4. Probabilities $p_{1},p_{3}$ and values
$B_{1},B_{3}$. Theorem 2}
\end{center}

It will be convenient to use the following technical result, which
allows instead of a simplex code to consider an orthogonal code.
Let
$\{\mbox{\boldmath $z$}_{i} \in {\mathbb{R}}^{n}, i=1,\ldots,M\}$
-- a simplex code with $\|\mbox{\boldmath $z$}_{i}\|^{2}= A$ and
$M \leq n$. Since
$\sum\limits_{k=1}^{M}\mbox{\boldmath $z$}_{k} = 0$, then denoting
$r = A/(M-1)$, we have
$$
\|\mbox{\boldmath $z$}_{i} - \mbox{\boldmath $z$}_{j}\|^{2} = 2rM;
\quad
(\mbox{\boldmath $z$}_{i},\mbox{\boldmath $z$}_{j}) =-r, \quad
i \neq j; \quad (\mbox{\boldmath $z$}_{3} - \mbox{\boldmath $z$}_{1},
\mbox{\boldmath $z$}_{3}- \mbox{\boldmath $z$}_{2}) = rM.
$$
Set an arbitrary vector
$\mbox{\boldmath $u$}_{0} \in {\mathbb{R}}^{n}$, such that
$\mbox{\boldmath $u$}_{0} \perp \{\mbox{\boldmath $z$}_{1}, \ldots,
\mbox{\boldmath $z$}_{M}\}$ and $\|\mbox{\boldmath $u$}_{0}\|^{2} = r$,
and consider vectors
$\mbox{\boldmath $u$}_{i} = \mbox{\boldmath $z$}_{i} +
\mbox{\boldmath $u$}_{0}$, $i = 1,\ldots,M$. Then
$(\mbox{\boldmath $u$}_{i},\mbox{\boldmath $u$}_{j}) = 0$,
$i \neq j$ and
$\|\mbox{\boldmath $z$}_{i} - \mbox{\boldmath $z$}_{j}\| =
\|\mbox{\boldmath $u$}_{i} - \mbox{\boldmath $u$}_{j}\|$ for any
$i,j$. In particular, we have
$\mbox{\boldmath $u$}_{0} = M^{-1}
\sum\limits_{j=1}^{M}\mbox{\boldmath $u$}_{j}$. This result can be
formulated as follows.

{P r o p o s i t i o n \,1}. {\it Let
$\{\mbox{\boldmath $z$}_{i} \in {\mathbb{R}}^{n}, i=1,\ldots,M\}$
be a simplex (i.e. equidistant) code with
$\|\mbox{\boldmath $z$}_{i}\|^{2}= A$. Then it can be represented as
\begin{equation}\label{simp1}
\begin{gathered}
\mbox{\boldmath $z$}_{i} = \mbox{\boldmath $u$}_{i} -
\mbox{\boldmath $u$}_{0}, \quad i = 1,\ldots,M, \qquad
\mbox{\boldmath $u$}_{0} = \frac{1}{M}
\sum_{j=1}^{M}\mbox{\boldmath $u$}_{j},
\end{gathered}
\end{equation}
where $\{\mbox{\boldmath $u$}_{i},  i=1,\ldots,M\}$ are mutually
orthogonal}
({\it i.e. $(\mbox{\boldmath $u$}_{i},\mbox{\boldmath $u$}_{j}) = 0$
for $i \neq j$}) {\it vectors with}
$\|\mbox{\boldmath $u$}_{i}\|^{2} = AM/(M-1)$.

Using Proposition 1 we replace vectors
$\{\mbox{\boldmath $x$}_{i}'\}$ by orthogonal vectors
$\{\mbox{\boldmath $u$}_{i} = \mbox{\boldmath $x$}_{i}' +
\mbox{\boldmath $u$}_{0}\}$ such that
$\|\mbox{\boldmath $u$}_{i}\|^{2} = A_{3}$ and
$(\mbox{\boldmath $u$}_{i},\mbox{\boldmath $u$}_{j}) = 0$,
$i \neq j$. Then $\mbox{\boldmath $x$}_{1}' -
\mbox{\boldmath $x$}_{i}' = \mbox{\boldmath $u$}_{1} -
\mbox{\boldmath $u$}_{i}$.
Denote
$$
\begin{gathered}
(\mbox{\boldmath $u$}_{i},\mbox{\boldmath $\xi$}') =
\sqrt{A_{3}}\,\xi_{i} = u_{i}A_{3}, \qquad
(\mbox{\boldmath $u$}_{i},\mbox{\boldmath $\eta$}') =
\sqrt{A_{3}}\,\eta_{i} = v_{i}A_{3}, \qquad i=1,\ldots,M.
\end{gathered}
$$

Note that if we would omit $\{\ln p_{i}\}$ from $\{b_{i}\}$, then,
for example, we have
$$
\begin{gathered}
B_{1} = -\ln {\mathbf E}_{\mbox{\boldmath $\xi$}'}e^{-b_{1}} \sim
\min_{u_{1},u_{2}}\left\{\frac{1}{4A_{4}}\left[A_{3} + A_{4} +
u_{1}\sqrt{A_{3}} - u_{2}\sqrt{A_{3}}\right]^{2} +
\frac{u_{1}^{2} + u_{2}^{2}}{2}\right\} = \\
= \frac{A_{3}(1+\beta)}{4},
\end{gathered}
$$
which corresponds to no-feedback case. Similar estimates would hold
for $B_{3}$ as well. Therefore for given
$\mbox{\boldmath $y$}' = \{u_{i}\}$ we should evaluate and take into
account conditional probabilities
$p_{k} = {\mathbf P}({\cal Z}_{k}\big|\mbox{\boldmath $y$}',
\mbox{\boldmath $x$}_{1}')$.

Note also that for large $M$ values $b_{1},b_{3},b_{4}$ are
approximately equal. Then it would be sufficient to evaluate how
close is $p_{2}$ to $1$.

In order to evaluate $p_{1}$, introduce events
$$
\begin{gathered}
{\cal A}_{12} = \left\{\max\{d_{1}^{(t)},d_{2}^{(t)}\} \leq
\min_{i \geq 3}d_{i}^{(t)} \leq \max\{d_{1}^{(t)},d_{2}^{(t)}\} +
\tau_{0}A_{3}\right\} = \\
= \left\{(d_{2}^{(t)}-d_{1}^{(t)})_{+} \leq \min_{i \geq 3}
d_{i}^{(t)} - d_{1}^{(t)} \leq
(d_{2}^{(t)}-d_{1}^{(t)})_{+} + \tau_{0}A_{3}\right\}, \\
{\cal A}_{13} = \left\{\max\{d_{1}^{(t)},d_{3}^{(t)}\} \leq
d_{2}^{(t)} \leq \max\{d_{1}^{(t)},d_{3}^{(t)}\} + \tau_{0}A_{3}
\right\}, \\
{\cal A}_{23} = \left\{\max\{d_{2}^{(t)},d_{3}^{(t)}\} \leq
d_{1}^{(t)} \leq \max\{d_{2}^{(t)},d_{3}^{(t)}\} + \tau_{0}A_{3}
\right\}.
\end{gathered}
$$
Then
\begin{equation}\label{P1u1u2}
\begin{gathered}
p_{1} \leq (M-1)^{2}[{\mathbf P}\left\{{\cal A}_{12}
|\mbox{\boldmath $y$}',\mbox{\boldmath$x$}_{1}'\right\} +
{\mathbf P}\left\{{\cal A}_{23}
|\mbox{\boldmath $y$}',\mbox{\boldmath $x$}_{1}'\right\} +
{\mathbf P}\left\{{\cal A}_{13}
|\mbox{\boldmath $y$}',\mbox{\boldmath $x$}_{1}'\right\}] \leq \\
\leq 3(M-1)^{2}{\mathbf P}\left\{{\cal A}_{12}
|\mbox{\boldmath $y$}',\mbox{\boldmath$x$}_{1}'\right\},
\end{gathered}
\end{equation}
since ${\mathbf P}\left\{{\cal A}_{23}
|\mbox{\boldmath $y$}',\mbox{\boldmath $x$}_{1}'\right\} \leq
\min\left[{\mathbf P}\left\{{\cal A}_{12}
|\mbox{\boldmath $y$}',\mbox{\boldmath$x$}_{1}'\right\},
{\mathbf P}\left\{{\cal A}_{13}
|\mbox{\boldmath $y$}',\mbox{\boldmath $x$}_{1}'\right\}\right]$,
and due to symmetry \\ ${\mathbf P}\left\{{\cal A}_{12}
|\mbox{\boldmath $y$}',\mbox{\boldmath$x$}_{1}'\right\} \sim
{\mathbf P}\left\{{\cal A}_{13}
|\mbox{\boldmath $y$}',\mbox{\boldmath $x$}_{1}'\right\}$. Therefore
it is sufficient to evaluate the probability \\
${\mathbf P}\left\{{\cal A}_{12}
|\mbox{\boldmath $y$}',\mbox{\boldmath$x$}_{1}'\right\}$.

If $\mbox{\boldmath $x$}_{\rm true}' =  \mbox{\boldmath $x$}_{1}'$,
then for $i \geq 2$
$$
\begin{gathered}
d_{i}^{(t)} - d_{1}^{(t)} = 2(\mbox{\boldmath $u$}_{1} -
\mbox{\boldmath $u$}_{i},\mbox{\boldmath $\xi$}' + \sigma
\mbox{\boldmath $\eta$}') + 2A_{3} = 2A_{3}(1+u_{1} - u_{i}) +
2\sigma\sqrt{A_{3}}(\eta_{1} - \eta_{i}).
\end{gathered}
$$
Denote
\begin{equation}\label{denote1}
\begin{gathered}
w_{i}\sigma = (1+u_{1} - u_{i})\sqrt{A_{3}}, \quad i = 2,\ldots, M,
\qquad s\sigma = \tau_{0}\sqrt{A_{3}}/2.
\end{gathered}
\end{equation}
Then ($\{\eta_{i}\}$ are independent ${\cal N}(0,1)$--Gaussian
random variables)
$$
\begin{gathered}
{\mathbf P}\left\{{\cal A}_{12}|\mbox{\boldmath $y$}',
\mbox{\boldmath $x$}_{1}'\right\} = {\mathbf P}
\left\{0 \leq \eta_{1} - (w_{2} + \eta_{1} - \eta_{2})_{+} +
\min_{i \geq 3}\{w_{i} - \eta_{i}\} \leq s
|\mbox{\boldmath $y$}'\right\} = \\
= {\mathbf P}\left\{0 \leq \eta_{1} + \min_{i \geq 3}
\{w_{i} - \eta_{i}\} \leq s, w_{2} + \eta_{1} - \eta_{2} < 0
|\mbox{\boldmath $y$}'\right\} + \\
+ {\mathbf P}\left\{0 \leq \eta_{2} - w_{2} + \min_{i \geq 3}
\{w_{i} - \eta_{i}\}\leq s, w_{2} + \eta_{1} - \eta_{2} \geq 0
|\mbox{\boldmath $y$}'\right\} \leq \\
\leq {\mathbf P}\left\{\eta_{1} + \min_{i \geq 3}
\{w_{i} - \eta_{i}\} \leq s|\mbox{\boldmath $y$}'\right\} +
{\mathbf P}\left\{\eta_{2} - w_{2} + \min_{i \geq 3}
\{w_{i} - \eta_{i}\}\leq s |\mbox{\boldmath $y$}'\right\} \leq \\
\leq (M-2)\left[{\mathbf P}\left\{w_{3}+ \eta_{1} - \eta_{3} \leq s
\right\} + {\mathbf P}\left\{w_{3} - w_{2}+ \eta_{2} - \eta_{3}
\leq s\right\}\right] \leq \\
\leq 2(M-2)e^{-[w_{3}- (w_{2})_{+}-s]_{+}^{2}/4},
\end{gathered}
$$
where on the last step the inequality (\ref{Phi1}) was used.

Using (\ref{P1u1u2}), for $p_{1}$ we get
\begin{equation}\label{q1}
\begin{gathered}
\ln p_{1} \leq -[w_{3}- (w_{2})_{+}-s]_{+}^{2}/4 + \ln (6M^{3}).
\end{gathered}
\end{equation}

Consider values $b_{1}$ and $B_{1}$. Below in brackets, for short,
we omit relatively small term $\ln(6M^{3})$, but it will be taken
into account in the final result. Using (\ref{q1}) we have
\begin{equation}\label{b1q11}
\begin{gathered}
b_{1} \geq \frac{1}{4A_{4}}\left[A_{3}+A_{4} -
(\mbox{\boldmath $u$}_{2} - \mbox{\boldmath $u$}_{1},
\mbox{\boldmath $\xi$}') + \frac{1}{4}(w_{3}-(w_{2})_{+} - s)_{+}^{2}
\right]_{+}^{2} = \\
= \frac{A_{3}}{4\beta}\left[1+ \beta + u_{1} - u_{2} +
\frac{1}{4A_{3}}(w_{3}-(w_{2})_{+} - s)_{+}^{2}\right]_{+}^{2} = \\
= \frac{A_{3}}{4\beta}\left[\beta+ y_{2} + \gamma
(y_{3}-(y_{2})_{+} - \tau_{0}/2)_{+}^{2}\right]_{+}^{2},
\end{gathered}
\end{equation}
where we denoted
\begin{equation}\label{defgamma}
\begin{gathered}
\gamma = 1/(4\sigma^{2}), \qquad
y_{i} = \sigma w_{i}/\sqrt{A_{3}} = 1 + u_{1} - u_{i}.
\end{gathered}
\end{equation}
Therefore (if integration limits are not pointed out then it is done
over all possible area)
\begin{equation}\label{B1q1}
\begin{gathered}
\left(\frac{2\pi}{A_{3}}\right)^{3/2}
{\mathbf E}_{\mbox{\boldmath $\xi$}'}e^{-b_{1}} =
\iiint\limits \exp\left\{-b_{1}(y_{2},y_{3}) -
\frac{A_{3}(u_{1}^{2} + u_{2}^{2} + u_{3}^{2})}{2}\right\}
du_{1}du_{2}du_{3} = \\
= \iiint\limits \exp\left\{-b_{1}(y_{2},y_{3}) - \frac{A_{3}}{2}
\left[u_{1}^{2}+ \left(1+u_{1} -y_{2}\right)^{2} + \left(1 +
u_{1} - y_{3}\right)^{2}\right]\right\}du_{1}dy_{2}dy_{3} = \\
= \sqrt{\frac{2\pi}{3A_{3}}}
\iint\limits e^{-b_{1}(y_{2},y_{3}) -A_{3}g(y_{2},y_{3})/3}
= \sqrt{\frac{2\pi}{3A_{3}}}
\iint\limits e^{-A_{3}f_{1}(y_{2},y_{3})/(12\beta)}
dy_{2}dy_{3},
\end{gathered}
\end{equation}
where
\begin{equation}\label{defg}
\begin{gathered}
g(u,v) = 1+u^{2}+v^{2}- uv - u-v =
\left(v - \frac{1+u}{2}\right)^{2} + \frac{3(1-u)^{2}}{4}, \\
f_{1}(y_{2},y_{3}) = 3\left[\beta+ y_{2} + \gamma
(y_{3}-(y_{2})_{+} - \tau_{0}/2)_{+}^{2}\right]_{+}^{2}+
4\beta g(y_{2},y_{3}).
\end{gathered}
\end{equation}
Represent the last integral in the right-hand side of (\ref{B1q1})
as follows
\begin{equation}\label{B1q11}
\begin{gathered}
\iint\limits e^{-A_{3}f_{1}(y_{2},y_{3})/(12\beta)}dy_{2}dy_{3} =
I_{1} + I_{2} + I_{3},
\end{gathered}
\end{equation}
where
\begin{equation}\label{B1q12}
\begin{gathered}
I_{i} = \iint\limits_{V_{i}}e^{-A_{3}f_{1}(y_{2},y_{3})/(12\beta)}
dy_{2}dy_{3}, \qquad i = 1,2,3,
\end{gathered}
\end{equation}
and
\begin{equation}\label{region11}
\begin{gathered}
V_{1} = \{\beta + y_{2} \leq 0\}, \qquad
V_{2} = \{-\beta \leq y_{2} \leq 0\}, \qquad
V_{3} = \{y_{2} \geq 0\}.
\end{gathered}
\end{equation}

We evaluate consecutively integrals $I_{1},I_{2},I_{3}$. For $I_{1}$
we have
\begin{equation}\label{B1q13}
\begin{gathered}
I_{1} \leq \iint\limits_{V_{1}}e^{-A_{3}g(y_{2},y_{3})/3}
dy_{2}dy_{3} = \\
= \int\limits_{y_{2} \leq -\beta}e^{-A_{3}(y_{2}-1)^{2}/4}
\int\limits_{-\infty}^{\infty}
e^{-A_{3}(2y_{3}-y_{2}-1)^{2}/12}dy_{3}dy_{2}
\sim e^{-A_{3}(1+\beta)^{2}/4}.
\end{gathered}
\end{equation}

Consider the integral $I_{2}$. Denoting
$z = \gamma(y_{3} - \tau_{0}/2)_{+}^{2}$,  we have
$$
\begin{gathered}
f_{1}(y_{2},y_{3}) = 3(y_{2} + \beta + z)^{2} +
4\beta g(y_{2},y_{3}) = \\
= (3+4\beta)\left[y_{2} + \frac{3z + \beta(1 -2y_{3})}{3+4\beta}
\right]^{2} + 3\beta(1 +\beta) + f_{3}(y_{3}), \\
f_{3}(y_{3}) = \frac{3\beta}{(3+4\beta)}\left\{
(1+\beta)(1-2y_{3})^{2} + 4z(z+1 +2\beta +y_{3})\right\}.
\end{gathered}
$$
Therefore
$$
\begin{gathered}
I_{2} \leq \iint\limits_{y_{2} \geq -\beta}
e^{-A_{3}f_{1}(y_{2},y_{3})/(12\beta)}dy_{2}dy_{3} \lesssim
e^{-A_{3}(1+\beta)/4}\int\limits_{-\infty}^{\infty}
e^{-A_{3}f_{3}(y_{3})/(12\beta)}dy_{3}.
\end{gathered}
$$
If $y_{3} \leq \tau_{0}/2$, then $z = 0$. Set some level
$\tau_{0}/2 < u < 1/2$. Then denoting
$z_{0} = \gamma(u - \tau_{0}/2)^{2}$, we have
$$
\begin{gathered}
\int\limits_{-\infty}^{\infty}e^{-A_{3}f_{3}(y_{3})/(12\beta)}dy_{3}
\leq \int\limits_{-\infty}^{u}\exp\left\{-\frac{A_{3}(1+\beta)}
{4(3+4\beta)}(1 -2y_{3})^{2}\right\}dy_{3} + \\
+ \int\limits_{u}^{\infty}\exp\left\{-\frac{A_{3}(1+\beta)}
{4(3+4\beta)}(1 -2y_{3})^{2} - \frac{A_{3}z_{0}
(z_{0}+1 +2\beta +u)}{(3+4\beta)}\right\}dy_{3} \lesssim \\
\lesssim \exp\left\{-\frac{A_{3}(1+\beta)}{4(3+4\beta)}(1 -2u)^{2}
\right\} + \exp\left\{- \frac{A_{3}z_{0}(z_{0}+1 +2\beta +u)}
{(3+4\beta)}\right\}.
\end{gathered}
$$
Set $u$, such that $(1+\beta)(1-2u)^{2} = 4z_{0}(1+2\beta)$, i.e.
set
$$
2u = \frac{\sqrt{1+\beta} + \tau_{0}\sqrt{\gamma(1+2\beta)}}
{\sqrt{1+\beta} + \sqrt{\gamma(1+2\beta)}}.
$$
Then we get
$$
\begin{gathered}
\int\limits_{-\infty}^{\infty}e^{-A_{3}f_{3}(y_{3})/(12\beta)}dy_{3}
\lesssim \exp\left\{-\frac{A_{3}(1+\beta)(1-\tau_{0})^{2}}
{4(3+4\beta)(1+2\sigma)^{2}}\right\}
\end{gathered}
$$
and therefore
\begin{equation}\label{B1q14}
\begin{gathered}
I_{2} \lesssim \exp\left\{-\frac{A_{3}(1+\beta)}{4}
\left[1 + \frac{(1-\tau_{0})^{2}}
{(3+4\beta)(1+2\sigma)^{2}}\right]\right\}.
\end{gathered}
\end{equation}

Consider the integral $I_{3}$ from (\ref{B1q12}). Represent it as
follows
\begin{equation}\label{I312}
\begin{gathered}
I_{3} = I_{31} + I_{32}, \\
I_{31} = \iint\limits_{y_{2} \geq 0,y_{3} \leq y_{2} + \tau_{0}/2}
e^{-A_{3}f_{1}(y_{2},y_{3})/(12\beta)}dy_{2}dy_{3}, \\
I_{32} = \iint\limits_{y_{2} \geq 0,y_{3} \geq y_{2}+ \tau_{0}/2}
e^{-A_{3}f_{1}(y_{2},y_{3})/(12\beta)}dy_{2}dy_{3}.
\end{gathered}
\end{equation}
If $y_{3} \leq y_{2} +\tau_{0}/2$, then
$$
\begin{gathered}
f_{1}(y_{2},y_{3}) = \beta\left(2y_{3} -1-y_{2}\right)^{2} +
3(1+\beta)(\beta + y_{2}^{2}).
\end{gathered}
$$
Integrating first over $y_{3}$, and then over $y_{2}$, we get
\begin{equation}\label{I31}
\begin{gathered}
I_{31} \lesssim e^{-A_{3}(1+\beta)/4}\int\limits_{0}^{\infty}
\exp\left\{-\frac{A_{3}}{12}(1-y_{2} - \tau_{0})_{+}^{2} -
\frac{A_{3}}{4\beta}(1+\beta)y_{2}^{2}\right\}dy_{2} \leq \\
\leq e^{-A_{3}(1+\beta)/4}\int\limits_{0}^{\infty}
\exp\left\{-\frac{A_{3}(3+4\beta)}{12\beta}\left[y_{2} -
\frac{\beta(1-\tau_{0})}{3+4\beta}\right]^{2} -
\frac{A_{3}(1+\beta)(1-\tau_{0})^{2}}{4(3+4\beta)}\right\}dy_{2} + \\
+ e^{-A_{3}(1+\beta)/4}\int\limits_{1-\tau_{0}}^{\infty}
e^{-A_{3}(1+\beta)y_{2}^{2}/(4\beta)}dy_{2}
\lesssim \exp\left\{-\frac{A_{3}(1+\beta)}{4}\left[1 +
\frac{(1-\tau_{0})^{2}}{3+4\beta}\right]\right\}.
\end{gathered}
\end{equation}

Consider the integral $I_{32}$. Denoting
$u = y_{3} - y_{2} - \tau_{0}/2$, we have
$$
\begin{gathered}
f_{1}(y_{2},y_{3}) \geq f_{4}(y_{2},u) = 3(\beta+ y_{2})^{2} +
6\gamma (\beta + y_{2})u^{2} + \\
+ 4\beta\left[1 +y_{2}^{2}+(u + \tau_{0}/2)^{2} +y_{2}(u + \tau_{0}/2)
- 2y_{2}- u - \tau_{0}/2\right] = \\
= u^{2}[6\gamma (\beta + y_{2}) + 4\beta]
- 4\beta u(1-\tau_{0} - y_{2}) + \\
+ 3(\beta + y_{2})^{2} + \beta(4+4y_{2}^{2} + \tau_{0}^{2} +
2y_{2}\tau_{0}-8y_{2}-2\tau_{0}).
\end{gathered}
$$
First we integrate over $u$ and then over $y_{2}$. Since
$$
\begin{gathered}
f_{4}(y_{2},0)= \left[y_{2}\sqrt{3+4\beta}- \frac{\beta(1-\tau_{0})}
{\sqrt{3+4\beta}}\right]^{2} - \frac{\beta^{2}(1-\tau_{0})^{2}}
{3+4\beta} + 3\beta^{2} + \beta(4-2\tau_{0} + \tau_{0}^{2}),
\end{gathered}
$$
we have
$$
\begin{gathered}
I_{32} \leq \iint\limits_{u \geq 0,y_{2} \geq 0}
e^{-A_{3}f_{4}(y_{2},u)/(12\beta)}dudy_{2} \lesssim \\
\lesssim e^{A_{3}(1-\tau_{0})^{2}/[6(3\gamma+2)]}
\int\limits_{y_{2} \geq 0}e^{-A_{3}f_{4}(y_{2},0)/(12\beta)}dy_{2}
\lesssim \\
\lesssim \exp\left\{-\frac{A_{3}(1+\beta)}{4}
\left[1 + \frac{(1-\tau_{0})^{2}}{3 + 4\beta} -
\frac{2(1-\tau_{0})^{2}}{3(1+\beta)(3\gamma+2)}\right]\right\}.
\end{gathered}
$$
then
\begin{equation}\label{I3}
\begin{gathered}
I_{3} \lesssim \exp\left\{-\frac{A_{3}(1+\beta)}{4}
\left[1 + \frac{(1-\tau_{0})^{2}}{3 + 4\beta} -
\frac{2(1-\tau_{0})^{2}}{3(1+\beta)(3\gamma+2)}\right]\right\}.
\end{gathered}
\end{equation}
That estimate is applicable for all $\gamma > 0$. If $\gamma$ is
small, then $\beta$ should be chosen such that
$\beta < 9\gamma/(2-9\gamma)$.

As a result, from (\ref{B1q1}), (\ref{B1q11}), (\ref{B1q13}),
(\ref{B1q14}) and (\ref{I3}) we get
\begin{equation}\label{B1q1a}
\begin{gathered}
{\mathbf E}_{\mbox{\boldmath $\xi$}'}e^{-b_{1}} \lesssim
e^{-A_{3}(1+\beta)^{2}/4} + \exp\left\{-\frac{A_{3}(1+\beta)}{4}
\left[1 + \frac{(1-\tau_{0})^{2}}
{(3+4\beta)(1+2\sigma)^{2}}\right]\right\} + \\
+ \exp\left\{-\frac{A_{3}(1+\beta)}{4}
\left[1 + \frac{(1-\tau_{0})^{2}}{3 + 4\beta} -
\frac{2(1-\tau_{0})^{2}}{3(1+\beta)(3\gamma+2)}\right]\right\}
\end{gathered}
\end{equation}
and then
\begin{equation}\label{B1a}
\begin{gathered}
B_{1} \geq \frac{A_{3}(1+\beta)}{4n} \times \\
\times \left[1 + \min\left\{\beta,\frac{(1-\tau_{0})^{2}}
{(3+4\beta)(1+2\sigma)^{2}},\frac{(1-\tau_{0})^{2}}{3 + 4\beta} -
\frac{2(1-\tau_{0})^{2}}{3(1+\beta)(3\gamma+2)}
\right\}\right] - \frac{3\ln M}{n},
\end{gathered}
\end{equation}
where the last term in the right-hand side of (\ref{B1a}) takes into
account the term omitted in (\ref{b1q11}).

Values $p_{3}$ and $B_{3}$ are evaluated similarly to values
$p_{1},B_{1}$. Introduce sets
$$
\begin{gathered}
{\cal Z}_{31} = \{\mbox{\boldmath $z$}': d_{23}^{(t)} > \tau_{0}A_{3},
\{\mbox{\boldmath $x$}'^{(1)t},\mbox{\boldmath $x$}'^{(2)t}\} =
\{\mbox{\boldmath $x$}_{1}',\mbox{\boldmath $x$}_{3}'\}\}, \\
{\cal Z}_{32} = \{\mbox{\boldmath $z$}': d_{23}^{(t)} > \tau_{0}A_{3},
\{\mbox{\boldmath $x$}'^{(1)t},\mbox{\boldmath $x$}'^{(2)t}\} =
\{\mbox{\boldmath $x$}_{2}',\mbox{\boldmath $x$}_{3}'\}\}
\end{gathered}
$$
and consider conditional probabilities
$P_{31} = {\mathbf P}\left\{{\cal Z}_{31}
|\mbox{\boldmath $y$}',\mbox{\boldmath $x$}_{1}'\right\}$ and
$P_{32} = {\mathbf P}\left\{{\cal Z}_{32}
|\mbox{\boldmath $y$}',\mbox{\boldmath $x$}_{1}'\right\}$.
Then
$$
\begin{gathered}
p_{3} \leq (M-2)\left(P_{31} + P_{32}\right) \leq 2(M-2)P_{31},
\end{gathered}
$$
since $P_{31} \geq P_{32}$. Then it is sufficient to evaluate
$P_{31}$. Using notations (\ref{denote1}), we have
$$
\begin{gathered}
P_{31} = {\mathbf P}\left\{\min_{i=2,4,5,\ldots}d_{i}^{(t)} >
\max\{d_{1}^{(t)},d_{3}^{(t)}\} + \tau_{0}A_{3}
|\mbox{\boldmath $y$}',\mbox{\boldmath $x$}_{1}'\right\} = \\
= {\mathbf P}\left\{\min_{i=2,4,5,\ldots}[w_{i} + \eta_{1} -
\eta_{i}] > (w_{3} + \eta_{1} - \eta_{3})_{+} + s
|\mbox{\boldmath $y$}'\right\} \leq \\
\leq {\mathbf P}\left\{w_{2} + \eta_{1} - \eta_{2} >
(w_{3} + \eta_{1} - \eta_{3})_{+} + s
|\mbox{\boldmath $y$}'\right\} = q_{1} + q_{2},
\end{gathered}
$$
where
$$
\begin{gathered}
q_{1} = {\mathbf P}\left\{\eta_{1} - \eta_{2} > s - w_{2},
\eta_{3} - \eta_{1} > w_{3}\right\}, \\
q_{2} = {\mathbf P}\left\{\eta_{3} - \eta_{2} > s- w_{2} + w_{3},
\eta_{1} - \eta_{3} > -w_{3}\right\}.
\end{gathered}
$$
Here, for example, $\eta_{1} - \eta_{2} \sim {\cal N}(0,1)$. For
probabilities $q_{1},q_{2}$ we use simple estimates
(see (\ref{Phi1}))
$$
\begin{gathered}
\ln q_{1} \leq -\frac{1}{4}\left[\max\left\{s-w_{2},w_{3}\right\}
\right]_{+}^{2}, \qquad \ln q_{2} \leq -\frac{1}{4}\left[
\max\left\{s-w_{2}+w_{3},-w_{3}\right\}\right]_{+}^{2}.
\end{gathered}
$$
Those estimates turn out to be sufficiently accurate, although it is
possible to strengthen them using dependence among random variables.
Then
$$
\begin{gathered}
\ln (P_{31}/2) \leq - \frac{1}{4}\left\{\min[\max\{s- w_{2},w_{3}\},
\max\{s-w_{2}+ w_{3},-w_{3}\}]\right\}_{+}^{2}.
\end{gathered}
$$

Using notations (\ref{defgamma}), after standard analysis we get
\begin{equation}\label{p3y}
\begin{gathered}
\ln p_{3} - \ln(4M) \leq -\gamma A_{3} r^{2}(y_{2},y_{3},\tau_{0}),
\end{gathered}
\end{equation}
where
\begin{equation}\label{r3}
\begin{gathered}
r(y_{2},y_{3},\tau_{0}) =
\begin{cases}
\min\{\tau_{0}/2-y_{2},-y_{3}\}, & y_{2} \leq \tau_{0}/2,
y_{3} \leq 0, \\
y_{3}, & y_{2} \leq \tau_{0}/2, y_{3} \geq 0, \\
0, & y_{2} \geq \tau_{0}/2, y_{3} \leq y_{2} - \tau_{0}/2, \\
\tau_{0}/2-y_{2}+y_{3}, & y_{2} \geq \tau_{0}/2,
y_{3} \geq y_{2} - \tau_{0}/2.
\end{cases}
\end{gathered}
\end{equation}

We have
\begin{equation}\label{b3}
\begin{gathered}
b_{3} = \frac{1}{4A_{2}}\left[A_{3}+A_{2} -
(\mbox{\boldmath $u$}_{2}' - \mbox{\boldmath $u$}_{1}',
\mbox{\boldmath $\xi$}') - \ln p_{3}\right]_{+}^{2}
\geq \frac{A_{3}}{4\mu}\left[\mu + y_{2} + \gamma
r^{2}(y_{2},y_{3},\tau_{0})\right]_{+}^{2},
\end{gathered}
\end{equation}
where $(y_{2},y_{3}) \in {\mathbf{R}}^{2}$. In order to simplify
the right-hand side of (\ref{b3}), first we evaluate contribution
to $B_{3}$ of points $(y_{2},y_{3}) \in D_{0}$, where
$$
\begin{gathered}
D_{0} = \{y_{2},y_{3}: y_{2} \leq -\mu\}.
\end{gathered}
$$
Using simple inequality
$f_{3}(y_{2},y_{3}) \geq 4\mu g(y_{2},y_{3})$, and integrating first
over $y_{2}$, and then over $y_{3}$, we get
\begin{equation}\label{J21}
\begin{gathered}
J_{0} = \iint\limits_{D_{0}}e^{-A_{3}g(y_{2},y_{3})/3}dy_{2}dy_{3}
\lesssim \int\limits_{-\infty}^{\infty}
e^{-A_{3}(2y_{3}-1+\mu)^{2}/12-A_{3}(1+\mu)^{2}/4}dy_{3} \sim
e^{-A_{3}(1+\mu)^{2}/4}.
\end{gathered}
\end{equation}

For remaining points $(y_{2},y_{3}) \in {\mathbf{R}}^{2} \setminus
D_{0} = \{y_{2},y_{3}: y_{2} > -\mu\} = {\cal D}$ we have
\begin{equation}\label{b3a}
\begin{gathered}
b_{3} \geq \frac{A_{3}}{4\mu}\left[\mu + y_{2} + \gamma
r^{2}(y_{2},y_{3},\tau_{0})\right]^{2}.
\end{gathered}
\end{equation}
In order to use the formula (\ref{r3}) it is convenient to partition
the remaining integration area ${\cal D}$ on four parts
\begin{equation}\label{region1}
\begin{gathered}
{\cal D} = \sum\limits_{i=1}^{4}D_{i},
\end{gathered}
\end{equation}
where
\begin{equation}\label{region2}
\begin{gathered}
D_{1} = \{-\mu \leq y_{2} \leq \tau_{0}/2,y_{3} \leq 0\}, \qquad
D_{2} = \{-\mu \leq y_{2} \leq \tau_{0}/2, y_{3} \geq 0\}, \\
D_{3} = \{y_{2} \geq \tau_{0}/2, y_{3} \leq y_{2} - \tau_{0}/2\},
\qquad D_{4} = \{y_{2} \geq \tau_{0}/2,
y_{3} \geq y_{2} - \tau_{0}/2\}.
\end{gathered}
\end{equation}
Then similarly to (\ref{B1q1}) we have
\begin{equation}\label{B3}
\begin{gathered}
\frac{2\pi\sqrt{3}}{A_{3}}
{\mathbf E}_{\mbox{\boldmath $\xi$}'}e^{-b_{3}} \leq J_{0} +
\sum\limits_{i=0}^{4}J_{i},
\end{gathered}
\end{equation}
where
\begin{equation}\label{deff3}
\begin{gathered}
J_{i} = \iint\limits_{D_{i}}e^{-A_{3}f_{3}(y_{2},y_{3})/(12\mu)}
dy_{2}dy_{3}, \quad i = 1,\ldots,4, \\
f_{3}(y_{2},y_{3}) = 3\left[\mu + y_{2} + \gamma
r^{2}(y_{2},y_{3},\tau_{0})\right]^{2} + 4\mu g(y_{2},y_{3}),
\end{gathered}
\end{equation}
and $g(u,v)$ is defined in (\ref{defg}).

We evaluate consecutively integrals $J_{1},\ldots.J_{4}$, starting
with $J_{1}$. For $(y_{2},y_{3}) \in D_{1}$ we have
$$
\begin{gathered}
f_{3}(y_{2},y_{3}) \geq 3(\mu + y_{2})^{2} + \mu
\left(2y_{3} - 1-y_{2}\right)^{2} + 3\mu(1-y_{2})^{2}.
\end{gathered}
$$
Integrating first over $y_{3} \leq 0$, and then over all $y_{3}$,
we get
\begin{equation}\label{J1}
\begin{gathered}
J_{1} = \iint\limits_{D_{1}}e^{-A_{3}f_{3}(y_{2},y_{3})/(12\mu)}
dy_{3}dy_{2} \lesssim \\
\lesssim \int\limits_{-\infty}^{\infty}
\exp\left\{-\frac{A_{3}}{12\mu}\left[3(\mu + y_{2})^{2} +
\mu(1+y_{2})^{2}+ 3\mu(1-y_{2})^{2}\right]\right\}dy_{2} \lesssim \\
\lesssim \exp\left\{- \frac{A_{3}(1+\mu)}{4}\left[1 +
\frac{1}{3+4\mu}\right]\right\}.
\end{gathered}
\end{equation}

Consider the integral $J_{2}$. Then
$$
\begin{gathered}
f_{3}(y_{2},y_{3}) = 3(\mu + y_{2} + \gamma y_{3}^{2})^{2} +
\mu(2y_{2} - 1-y_{3})^{2} + 3\mu (1-y_{3})^{2} = \\
= \left[\sqrt{3+4\mu}y_{2} + \frac{\mu - 2\mu y_{3}+
3\gamma y_{3}^{2}}{\sqrt{3+4\mu}}\right]^{2} + f_{31}(y_{3}),
\end{gathered}
$$
where
$$
\begin{gathered}
f_{31}(y_{3}) = \frac{12\mu}{(3+4\mu)}\left\{(1+\mu)^{2} +
\gamma^{2}y_{3}^{4} + \gamma y_{3}^{2}(1+2\mu + y_{3}) +
(1+\mu)(y_{3}^{2}-y_{3})\right\} \geq \\
\geq \frac{12\mu(1+\mu)}{(3+4\mu)}\left\{1+\mu +
(\gamma+1)\left[y_{3} - \frac{1}{2(\gamma+1)}\right]^{2} -
\frac{1}{4(\gamma+1)}\right\}.
\end{gathered}
$$
Therefore integrating first over all $y_{2}$, and then over all
$y_{3}$, we get
\begin{equation}\label{J2}
\begin{gathered}
J_{2} \lesssim \exp\left\{- \frac{A_{3}(1+\mu)}{4}\left[1 +
\frac{1}{3+4\mu} - \frac{1}{(3+4\mu)(\gamma + 1)}\right]\right\}.
\end{gathered}
\end{equation}

For the integral $J_{3}$ similarly to (\ref{J1}) we get
\begin{equation}\label{J3}
\begin{gathered}
J_{3} \lesssim \exp\left\{- \frac{A_{3}(1+\mu)}{4}\left[1 +
\frac{1}{3+4\mu}\right]\right\}.
\end{gathered}
\end{equation}

Consider the integral $J_{4}$. Denoting
$z = \tau_{0}/2-y_{2}+y_{3}$, we have
$$
\begin{gathered}
f_{3}(y_{2},z) = (3+4\mu)\left[y_{2} + \frac{3\gamma z^{2} +
\mu(2z-\tau_{0}-1)}{3+4\mu}\right]^{2} + 3\mu(1+\mu) + f_{3}(z), \\
f_{3}(z) = \frac{3\mu}{(3+4\mu)}\left\{4\gamma^{2}z^{4} +
2\gamma z^{2}(4+4\mu +\tau_{0}-2z) +
(1+\mu)(2z-\tau_{0}-1)^{2}\right\}.
\end{gathered}
$$
Note that
$$
\begin{gathered}
3\gamma z^{2} + \mu(2z-\tau_{0}-1) \geq 0, \qquad z \geq z_{0} =
\frac{1+\tau_{0}}{1+ \sqrt{1+3\gamma(1+\tau_{0})/\mu}}.
\end{gathered}
$$
Therefore for $z \leq z_{0}$ we integrate over all $y_{2} \geq 0$
$$
\begin{gathered}
\iint\limits_{0 \leq z \leq z_{0},y_{2} \geq 0}
e^{-A_{3}f_{3}(y_{2},y_{3})/(12\mu)}dy_{2}dy_{3}
\lesssim e^{-A_{3}(1+\mu)/4}
\exp\left\{-\frac{A_{3}}{12\mu}\min_{0 \leq z \leq z_{0}}
f_{3}(z)\right\} \lesssim \\
\lesssim \exp\left\{-\frac{A_{3}(1+\mu)}{4}\left[1 +
\frac{(1+\tau_{0})^{2}}{(3+4\mu)}\left[1 -
\frac{2}{1+ \sqrt{1+3\gamma(1+\tau_{0})/\mu}}\right]^{2}\right]
\right\},
\end{gathered}
$$
since
$$
\begin{gathered}
\min_{0 \leq z \leq z_{0}}f_{3}(z) \geq
\frac{3\mu(1+\mu)(1+\tau_{0})^{2}}{(3+4\mu)}\left[1 -
\frac{2}{1+ \sqrt{1+3\gamma(1+\tau_{0})/\mu}}\right]^{2}.
\end{gathered}
$$

If $z \geq z_{0}$ then minimum of the function $f_{3}(y_{2},z)$ is
attained for $y_{2} = 0$. Also
$$
\begin{gathered}
f_{3}(0,z) \geq 3\mu(1+\mu) + 2\mu \left\{\left[\sqrt{3\gamma+2}z -
\frac{1+\tau_{0}}{\sqrt{3\gamma+2}}\right]^{2}\right\} +
\frac{3\mu\gamma (1+\tau_{0})^{2}}{3\gamma+2}.
\end{gathered}
$$
Therefore
$$
\begin{gathered}
\iint\limits_{z \geq z_{0},y_{2} \geq 0}
e^{-A_{3}f_{3}(y_{2},z)/(12\mu)}dy_{2}dz \leq
\int\limits_{z \geq z_{0}}e^{-A_{3}f_{3}(0,z)/(12\mu)}dz \lesssim \\
\lesssim \exp\left\{-\frac{A_{3}(1+\mu)}{4}\left[1 +
\frac{\gamma (1+\tau_{0})^{2}}{(3\gamma+2)(1+\mu)}\right]\right\},
\end{gathered}
$$
which gives
\begin{equation}\label{J4}
\begin{gathered}
\ln J_{4} \lesssim - \frac{A_{3}(1+\mu)}{4}\left[1 +
\frac{(1+\tau_{0})^{2}}{(3+4\mu)}\min\left\{
\frac{\gamma}{1+\gamma},\left[1 -
\frac{2}{1+ \sqrt{1+3\gamma(1+\tau_{0})/\mu}}\right]^{2}\right\}
\right].
\end{gathered}
\end{equation}

Then from (\ref{B3}), (\ref{J21}) and (\ref{J1})--(\ref{J4}) we get
\begin{equation}\label{B3a}
\begin{gathered}
B_{3} \gtrsim \frac{A_{3}(1+\mu)}{4n}\left[1 +
\min\left\{\mu, \frac{\gamma}{(3+4\mu)(1+\gamma)}\left[1 -
\frac{2}{1+ \sqrt{1+3\gamma/\mu}}\right]^{2}\right\}\right].
\end{gathered}
\end{equation}

As a result, from (\ref{genFB}) we get a general result for any
$\sigma < \infty$.

{T h e o r e m \,2}. Let $\ln M = o(n)$, $n \to \infty$. Then for
any $\sigma < \infty$ the inequality holds
\begin{equation}\label{F1}
F(M,A,\sigma) \geq F_{1}(M,A,\sigma) \geq \max_{\beta,\tau_{0}}
\min\limits_{k = 1,2,3}B_{k} + o(1) > E(M,A), \qquad n \to \infty,
\end{equation}
where values $B_{1},B_{2},B_{3}$ are defined in (\ref{B1a}),
(\ref{defe2}) and  (\ref{B3a}), respectively.

The relation (\ref{F1}) has been proved provided $M \leq (n+2)/2$.
In fact, the formula (\ref{F1}) remains valid for any $M$ such that
$M = e^{o(n)}$, $n \to \infty$.
Indeed, note that instead of simplex codes
$\{\mbox{\boldmath $x$}_{i}'\}$ or $\{\mbox{\boldmath $x$}_{i}''\}$
on phases I--II we may use ``almost'' equidistant codes, for which,
for example,
$\|\mbox{\boldmath $x$}_{i}' - \mbox{\boldmath $x$}_{j}'\|^{2} =
2An(1 + o(1))$, $n \to \infty$, $i \neq j$. All calculations then
remain essentially the same. Such codes do exist due to
the following result.

Denote by $S_{n}$ the unit sphere in $\mathbb{R}^{n}$
centered at ${\mathbf 0}$.

{P r o p o s i t i o n \,2}. {\it For any $\rho \in (0,1)$ and
$n \geq 3$ there exists a code ${\cal C} = \{\mbox{\boldmath $x$}_{1},
\ldots, \mbox{\boldmath $x$}_{M}\} \subset S_{n}$ with
$|(\mbox{\boldmath $x$}_{i},\mbox{\boldmath $x$}_{j})| \leq \rho$,
$i \neq j$, such that}
\begin{equation}\label{State2}
\begin{gathered}
M \geq \rho e^{n\rho^{2}/2}.
\end{gathered}
\end{equation}

{P r o o f}. Denote by $\Omega(\theta)$ the area of the ``cap'' cut
out from $S_{n}$ by the cone of half-angle $\theta$. In particular,
the area of $S_{n}$ equals $\Omega(\pi)$. Then for any
$0 < \theta < \pi/2$ there exists a code
${\cal C} = \{\mbox{\boldmath $x$}_{1},\ldots,
\mbox{\boldmath $x$}_{M}\} \subset S^{n}$ with
$|(\mbox{\boldmath $x$}_{i},\mbox{\boldmath $x$}_{j})| \leq
\cos \theta$, $i \neq j$, such that
$$
M \geq \frac{\Omega(\pi)}{2\Omega(\theta)}.
$$
For the ratio $\Omega(\theta)/\Omega(\pi)$ the following estimate
is known \cite[formula (27)]{Sh}
$$
\begin{gathered}
\frac{\Omega(\theta)}{\Omega(\pi)} \leq
\frac{\Gamma\left(\dfrac{n}{2}+1\right)\sin^{n-1} \theta}
{n\Gamma\left(\dfrac{n+1}{2}\right)\sqrt{\pi}\cos \theta}.
\end{gathered}
$$
Using that estimate for $\rho = \cos\theta$ we get
$$
\begin{gathered}
M \geq \frac{\Omega(\pi)}{2\Omega(\theta)} \geq
\frac{n\Gamma\left(\dfrac{n+1}{2}\right)\sqrt{\pi}\rho}
{2\Gamma\left(\dfrac{n}{2}+1\right)(1-\rho^{2})^{(n-1)/2}} \geq
\frac{n\Gamma\left(\dfrac{n+1}{2}\right)\sqrt{\pi}\rho e^{n\rho^{2}/2}}
{2\Gamma\left(\dfrac{n}{2}+1\right)\sqrt{e}}.
\end{gathered}
$$
From the last inequality for $n \geq 3$ the estimate (\ref{State2})
follows. $\qquad \triangle$

\begin{center}
{\large\bf \S\,5. Proof of Theorem 1}
\end{center}

We need to investigate asymptotics of values $B_{1}, B_{2}$ and
$B_{3}$ when $\sigma \to 0$ and $\sigma \to \infty$.

a) If $\sigma \to 0$ then set $\tau_{0} > 0$ such that
$\tau_{0} \to 0$. Then for $B_{1}$ from (\ref{B1a}) we have
as $n \to \infty$, $\sigma \to 0$
$$
\begin{gathered}
B_{1} \geq \frac{AM}{4(M-1)}\left[1 + \min\left\{\beta,
\frac{1}{3+4\beta}\right\}\right] + o(1).
\end{gathered}
$$

For $B_{2}$ from (\ref{defe2}) we have
$$
\begin{gathered}
B_{2} \geq \frac{AM}{4(M-1)}\left[1 + \frac{\beta}{1+\beta} -
\frac{2\beta}{(1+\beta)M}\right] + o(1).
\end{gathered}
$$

For $B_{3}$ from (\ref{B3a}) we have
$$
\begin{gathered}
B_{3} \geq \frac{AM(1+\mu)}{4(M-1)(1+\beta)}
\left[1 + \min\left\{\mu, \frac{1}{3+4\mu}\right\}\right] \geq \\
\geq \frac{AM}{4(M-1)}\left[1 + \min\left\{\beta, \frac{1}{3+4\beta}
\right\} -\frac{\beta}{M(1+\beta)}\right] + o(1).
\end{gathered}
$$

Then we get as $\sigma \to 0$ and $n \to \infty$
$$
\min\limits_{k = 1,2,3}B_{k} \geq \frac{AM}{4(M-1)}\left[1 +
\min\left\{\frac{\beta}{1+\beta},\frac{1}{3+4\beta}\right\} -
\frac{2\beta}{(1+\beta)M} + o(1)\right].
$$

We set $\beta$ such that both terms under minimization become equal,
i.e. set
$$
\beta = (\sqrt{5} - 1)/4 \approx 0.3090.
$$
Then we get
$$
\min\limits_{k = 1,2,3}B_{k} \geq \frac{AM}{4(M-1)}\left[
1 + \frac{1}{2+\sqrt{5}} - \frac{1}{2M} + o(1)\right],
$$
from which the formula (\ref{Theor1a}) follows.

b) In that case $\sigma \to \infty$, i.e. $\gamma \to 0$. Set
$\tau_{0} > 0$ such that $\tau_{0} \to 0$ and choose
$\beta = \gamma/7,1$. Then after simple calculations we get
$$
\min\limits_{k = 1,2,3}B_{k} \geq \frac{AM}{4(M-1)}\left[
1 + \frac{\gamma}{14} + o(\gamma^{2})\right] - \frac{3\ln M}{n},
\qquad \gamma \to 0,
$$
from which the formula (\ref{Theor1b}) follows.

Note that in both extreme case as $\sigma \to 0$ or
$\sigma \to \infty$ the value $\tau_{0}$ was chosen such that
$\tau_{0} > 0$, but $\tau_{0} \to 0$. For intermediate values of
$\sigma$ optimal $\tau_{0} \not\to 0$.

\medskip

The authors wish to thank  V.V. Prelov for useful discussions and
constructive critical remarks. They are also grateful to the
University of Tokyo for supporting this joint research.


\vspace{5mm}

\begin{flushleft}
{\small {\it Burnashev Marat Valievich} \\
Kharkevich Institute for Information Transmission Problems, \\
Russian Academy of Sciences, Moscow\\
 {\tt burn@iitp.ru}} \\
{\small {\it Yamamoto Hirosuke} \\
School of Frontier Sciences \\
The University of Tokyo, Japan \\
 {\tt hirosuke@ieee.org}}
\end{flushleft}%

\end{document}